# INFORMATION SECURITY IN HEALTH CARE CENTRE USING CRYPTOGRAPHY AND STEGANOGRAPHY

**BABATUNDE, A.O.[1], TAIWO, A. J.[1], DADA, E. G.[2]**

[1]Computer Science Department, University of Ilorin, Ilorin, Kwara State, Nigeria
[2]Computer Engineering Department, University of Maiduguri, Maiduguri, Nigeria

Corresponding Author: gbengadada@unimaid.edu.ng

**ABSTRACT**

As the volume of medicinal information stored electronically increase, so do the need to enhance how it is secured. The inaccessibility to patient record at the ideal time can prompt death toll and also well degrade the level of health care services rendered by the medicinal professionals. Criminal assaults in social insurance have expanded by 125% since 2010 and are now the leading cause of medical data breaches. This study therefore presents the combination of 3DES and LSB to improve security measure applied on medical data. Java programming language was used to develop a simulation program for the experiment. The result shows medical data can be stored, shared, and managed in a reliable and secure manner using the combined model.

## 1.0 INTRODUCTION

In health industries, storing, sharing and management of patient information have been influenced by the current technology. That is, medical centres employ electronical means to support their mode of service in order to deliver quality health services. The importance of the patient record cannot be over emphasised as it contributes to when, where, how, and how lives can be saved. About 91% of health care organizations have encountered no less than one data breach, costing more than $2 million on average per organization [1-3]. Report also shows that, medical records attract high degree of importance to hoodlums compare to Mastercard information because they infer more cash base on the fact that bank

cards can rapidly be crossed out in order to prevent the potential harm while medical data cannot be so easily destroyed [4].

According to [5], specialists concur that while encryption is fundamental, securing medicinal information ought not depend on this approach alone or on other specialized fixes, for example, antivirus projects and firewalls to shield the protection and security of information in health centres. The frail purpose of encryption is in fact misuse once a day by intruders. Therefore, there is need for more advance and sophisticated approach to secure information in a medical or health centre.

Both Cryptography and Steganography has been the major means of protecting data. Although, they have distinct goal which makes them to be identify as different method of securing data. Cryptography clouds the importance (that is, sense) of a message, however it doesn't hide the existence of the message while Steganography is concerned with concealing the existence of secret information within computer files [6]. Therefore, this study employs 3DES and LSB to form a combine cryptography and steganography technique to improve the security of medical data.

## 1.2 TRIPLE DATA ENCRYPTION STANDARD

Triple Data Encryption Standard (3DES) was made from DES calculation, developed in the mid-1970s utilizing 56-bit key. The powerful security 3DES gives is just 112 bits because of meet-in-the-centre assaults. Triple DES runs three times slower than DES, however is significantly more secure if utilized appropriately. The methodology for unscrambling something is the same as the technique for encryption, aside from it is executed backward. Triple DES calculation utilizes three emphases of normal DES figure. It gets a mystery 168-piece key, which is partitioned into three 56-bit keys [7]. The build-up of 3DES compasses of Encryption utilizing the primary mystery key, Decryption utilizing the second mystery key, Encryption utilizing the third mystery key.

## 1.3   LEAST SIGNIFICANCE BIT

Least Significant Bit (LSB) is one of the conventional strategies in picture steganography. It deals with the manipulation of the bit of that is least significant in the cover image to encode the secret information. It is known with the ability to accommodate huge amount of data [8].

## 2.0   RELATED WORK

Many researchers have worked on security of data using cryptography, steganography and combination of the two technique. Some researchers carried out the perfonmance analysis of the alogrithms. Some of the research performed on data or information security are; Sachin and Kumar [9], carried out a research on GSM network. In their research, they gave additional encryption to GSM network by implementing Data Encryption Standard (DES), Triple Data Encryption (3DES) and Advanced Data Encryption (AES). Also, the study carried out the performance analysis of the three algorithms against Brute Force attack implemented in MATLAB environment. It was discovered that 3DES is better in security performance than DES but not AES because it takes a longer time for the brute force attack to breach AES security than 3DES and DES, similarly the time taken by brute attack to breach 3DES security is more than DES.

Zodape and Shulka [10], presented the performance analysis of 3DES and RSA algorithm. 3DES and RSA was used to secure image steganography. In their result, it was revealed that 3DES takes less time, less complex, more secure and the changes in the image bit is less; hence 3DES is more effective in securing image steganography.

Ganorkar [11], worked on intergrity and security of data in cloud. The study implements 3DES Encryption algorithm to secure the data stored on cloud. It was concluded that 3DES is good for securing textual data.

Sandeep, Sachin and Naveen [12], worked on data security by combining cryptography and image steganography. The study used AES and Least Significant Bit(LSB) for cryptography and steganoography respectively to enhance security of data. They concluded that LSB is a very good technique for embeeding sensitve data behind a media carrier.

Punita, Kamal and Manish [13], carried out a research on performance evaluation of audio steganography. The study used Discrete Cosine Transform (DCT) and LSB to hide secrete audio file. DCT was used to compress the secrete audio file and LSB was used to encrypt the audio file and also embed the encrypted audio file behind an image carrier. The performance evaluation was measured using Peak Signal to Noise Ratio(PSNR) and Mean Square Error(MSE). It was discovered that LSB has a good perfoemance when used with gray scale platte image.

Thomas and Panchami [14], worked on the security of messages transferred through Short Message Service(SMS). The study used blowfish algorithm fro encryption and decryption of text message being sent through SMS from one mobile phone to another. It was concluded that blowfish alggorithm is fast and saves battery of the mobile phone.

Syeda and Durga [15] researched on securing data transmission using cryptography and steganography. The study implemented AES, Discrete Wavelet Transform (DWT) and LSB in securing data transmitted. AES as a cryptography algorithm was used to encrypt the data , DWT was applied to the cover image to get sub band image and LSB was used to embed the cipher text behind the sub band image. LSB was discovered to have a good performance when measured with MSE and PSNR.

In summary, all the related studies reviewd shows that steganography technique is not ment to replace cryptography as both technique have different advantage and disadvantage. It can be concluded that the limitation of one technique can be address by another technique. Therefore, this study combine cryptography technique with steganography in order to achieve a more robust and secure syatem for mediacal data.

### 3.0 METHODOLOGY

The method used in this research is the combination of cryptography and steganography. That is, 3DES algorithm for encryption described in figure 1 and LSB for steganography described in figure 2. The methodology employed entails two main steps to secure medical data in text format which stated as follows.

(a)　The first step will accept patient data in plain text format as input and encrypt the plain text using 3DES to generate cipher text as output. This step requires a key called secret key

(b)　The second step will accept the output of the first step as input and embed it in a cover image using LSB to generate an output in image form called stego image.

In order to retrieve data for the stego image, the steps involve are mainly two.

(a)　The first step in retrieving the secured data is to extract the cipher text form the stego image.

(b)　The second step to finally retrieve the secure data is to decrypt the cipher text using 3DES algorithm. This step requires a secret key which correspond to the key used during encryption. Figure 3.3, Figure 3.4 and Figure 3.5 shows the flow chart for encryption process, decryption process and the entire process of the proposed stegacrypt system.

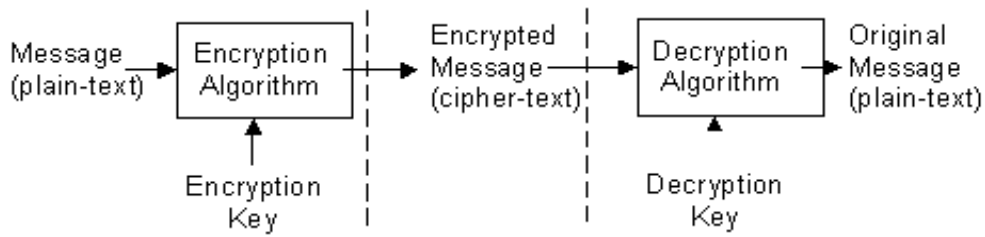

**Fig. 1:** The Encryption Scheme

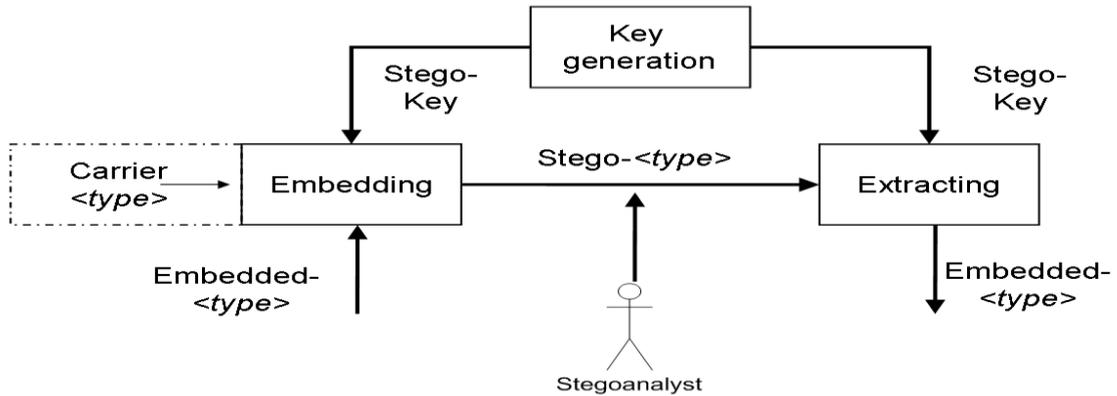

**Fig. 2:** Conceptual view of steganography scheme

## 4.0 RESULT AND DISCUSSION

**SECURE PHASE**

This is the phase that involves the securing process of patient health details. Figure 3 show the interface that allow then medical practitioners to select patient record to be secured and the cover image to be used to hide the data. The "Select File/Data" button and "Select Cover Image" button is use respectively for input selection.

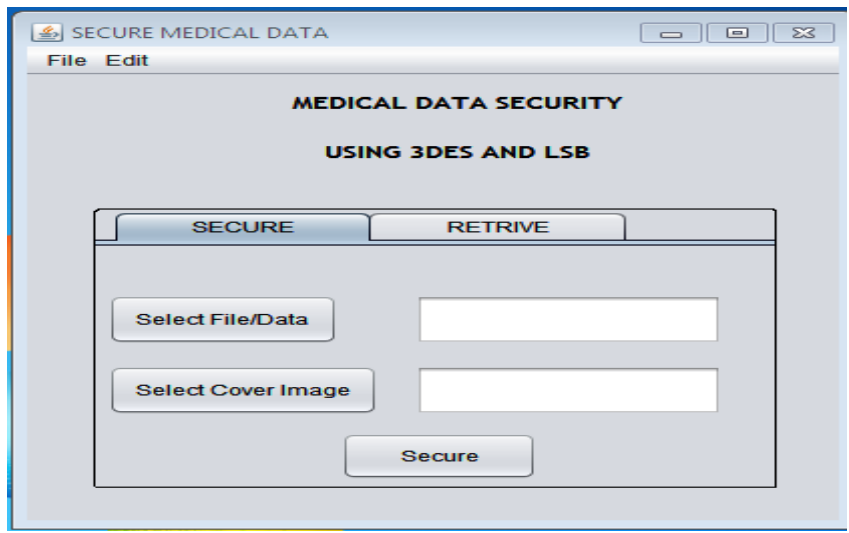

**Fig. 3:** The Secure interface

Figure 4 shows the Input selection in progress. The "open" button is clicked to complete the selection of a particular input.

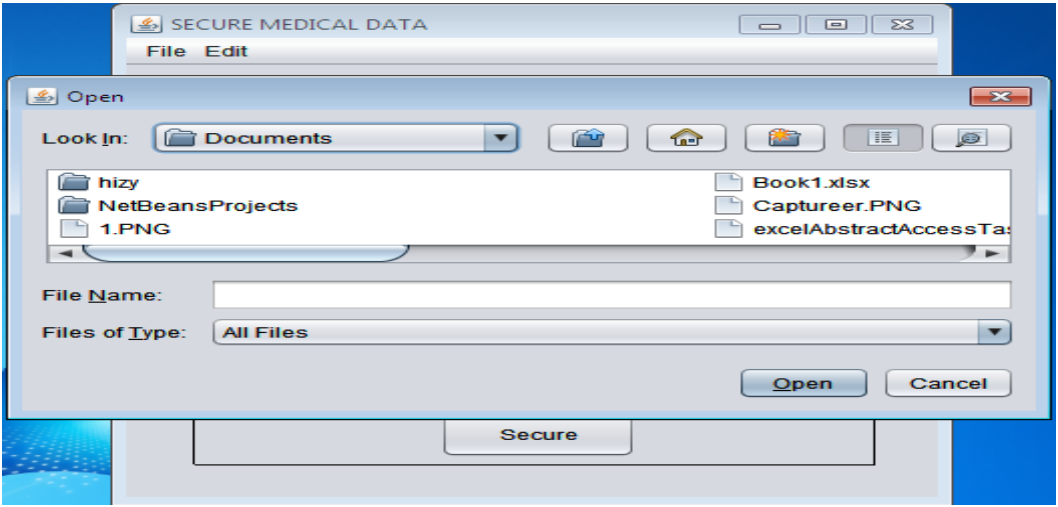

**Fig. 4:** The medical data selection process for securing operation

Figure 5 indicate the complete and sucessful selection of required input. The "Secure" button is then clicked to trigger the securing operation of the patient medical record. After a successful secure process, the output of this stage is a stego image.

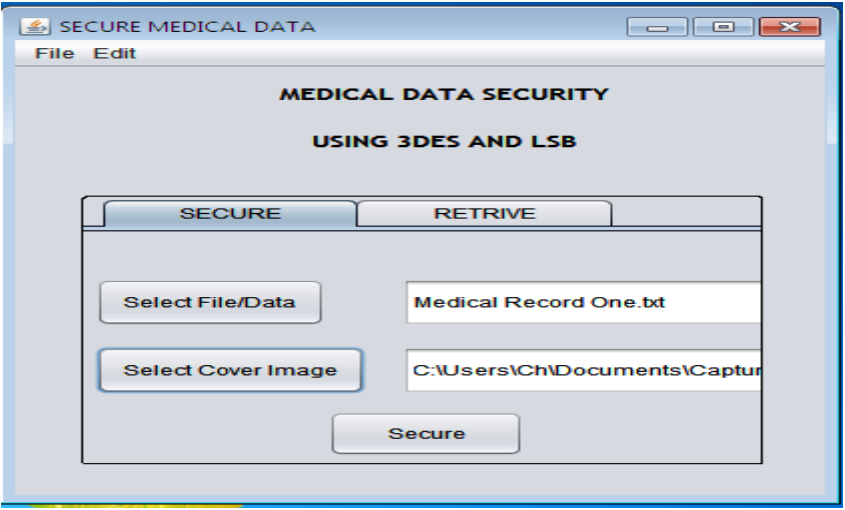

**Fig. 5:** Complete selection of Medical record and Cover Image.

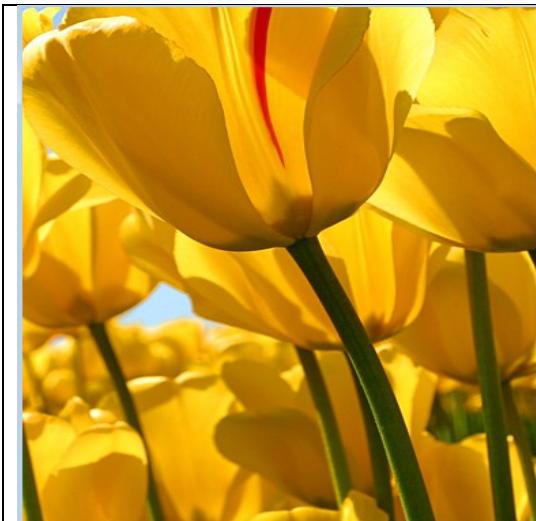 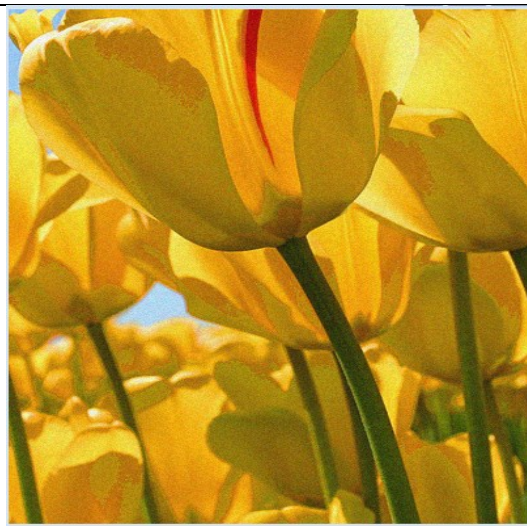

**Fig. 6:** Sample of Cover Image used   **Fig. 7:** Sample of Generated Stego Image

**RETRIEVE PHASE**

The retrieve phase is the reverse of the secure phase. It helps to get the secured medical data from the Stego image generated in secure phase. Figure 8 shows the retrieving interface. It contains two buttons namely, "Select Secured File/Data" and "Select coverImage" button which is used to select the required input.

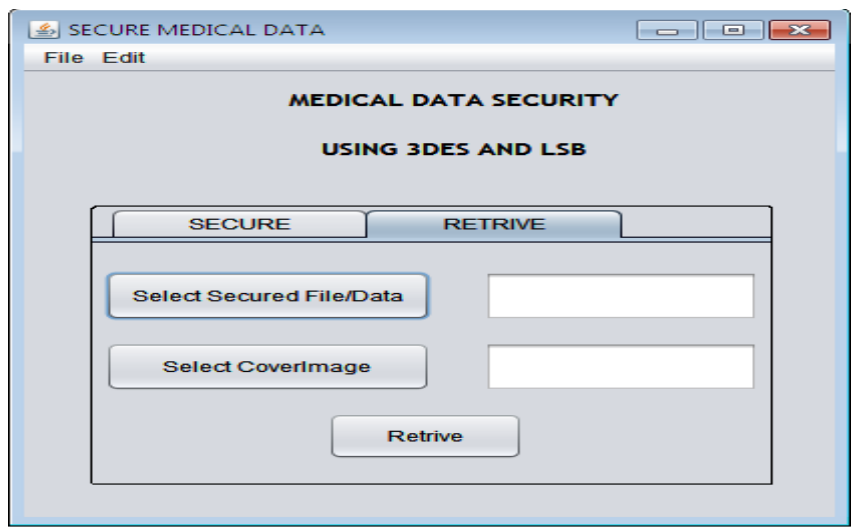

**Fig. 8:** The Retrieve Interface

Figure 9 shows the selection of required process in progress. The "open" button is clicked to complete the selection of a particular input.

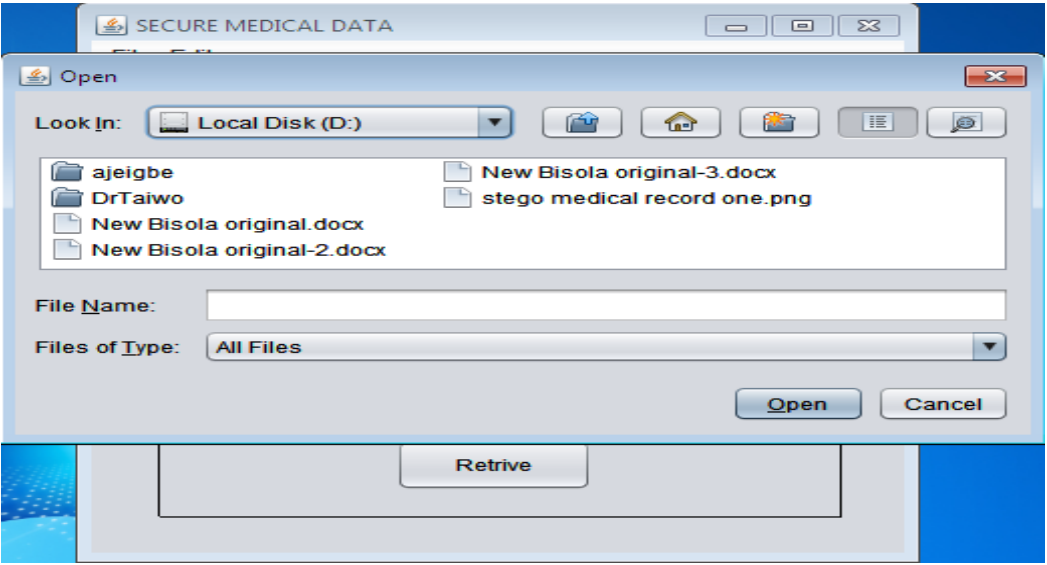

**Fig. 9:** The medical data selection process for retrieving operation

Figure 10 shows the complete selection of required input to retrieve the patient medical data. The "Retrieve" button is then clicked to trigger the retrieving process. After a successful retrieving process, the medical data secured is accessible and ready for use.

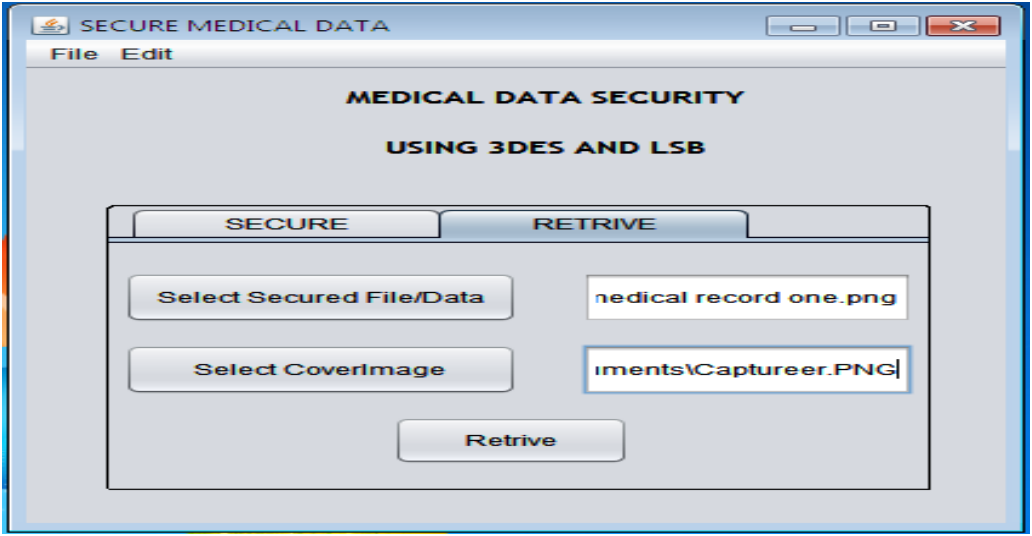

**Fig. 10:** Complete selection of the required input for retrieving process.

## 4.1 RESULT ANALYSIS

This study analyses the result obtained by comparing the combined or integrated method used in this study to each of the single technique. That is, 3DES and LSB. The combined method is found to be more secure, reliable as it offers double layer of security with different technique to medical data and require little or no technical knowledge. Table 1 shows the simple comparison between the combined method, 3DES and LSB.

**Table 1:** Simple comparison between the combined method, 3DES and LSB

|  | **Combined Method** | **3DES** | **LSB** |
|---|---|---|---|
| **Security Layers** | Double Layer | Single Layer | Single Layer |
| **Reliability** | 65% Reliable | 50% Reliable | 47% Reliable |
| **Speed** | 60% Fast | 75% Faster | 75% Faster |
| **Numbers of Keys** | Two (Secrete key & Cover Image) | Single (Secrete key) | Single (Cover Image) |
| **Differential Cryptanalytic Attack** | Not Applicable | Factor of 4 | Not Applicable |
| **Number of Rounds** | Not Applicable | 48 | Not Applicable |
| **Key Length in Byte** | Varies ($112^5$ and above) | $112^5$ | Not Applicable |

In conclusion, the combined method poses low operation speed as the only limitation while it surpasses 3DES and LSB in terms of Reliability strength of security.

## 5.0 SUMMARY and CONCLUSION

Among other sectors such as academic, banking and finance, medical sector is not an exception in experiencing cyber-attack as patient health data are also stored and shared electronically through the internet. The intrusion and stealing of patient data automatically affects the service offered at the health care centres and thus, poses threat to life of the patient. This study employs both cryptography and steganography technique to improve the security of medical data. Encryption and data hiding are the two main steps involved in the method use in this study to secure medical data while the reverse is use in opposite direction to retrieve the medical data. That is, 3DES is first use to covert patient data in to cipher text before it is hidden in an image carrier using LSB. The result obtained in the securing stage is called stego image which serves as an input to the retrieving stage. The retrieving process starts with LSB by extracting the scrambled data from the stego image and decrypt using 3DES. The analysis shows that the result obtains in this study correspond to the aim intended to achieve.

In conclusion, this study demonstrates a double layer security, that is, the combination of cryptography (3DES encryption algorithm) and steganography (LSB image encoding technique) to improve data security in medical or health care organization. The havoc caused by cyber-attack on data relating to health issues motivates this study to demonstrate a double layer security system. The result in this study shows a double layered security system offer more security strength and make successful cyber-attack to be more difficult. In fact, medical data can be hidden within another medical data. That is, medical data in image format with no secret information can be used as a cover for other medical data that need to be secured.

In future, a faster and secure technique should be investigated because health organisation is a very critical section that deals with life and therefore secured but too slow security technique might cause more havoc than the advantage it offers.

**REFERENCE**

[1]     UIC, (2017). Why Data Security is The Biggest Concern of Health Care. Retrieved from: http://healthinformatics.uic.edu/resources/articles/why-data-security-is-the-biggest-concern-of-health-care/ Retrieved date: 17th July, 2017.11